\documentclass[11pt,letterpaper]{article}
\usepackage[utf8]{inputenc}
\DeclareUnicodeCharacter{03B8}{\circ}
\usepackage{graphicx}
\usepackage{scalerel}
\usepackage{amsmath}
\usepackage{booktabs}
\usepackage{tabu}
\usepackage{graphicx}
\usepackage{caption}
\usepackage{subcaption}
\usepackage{cite}
\usepackage{secdot}
\usepackage[margin=0.8in]{geometry}
\usepackage{amssymb}
\usepackage{url}
\usepackage[colorlinks = true,
            linkcolor = blue,
            urlcolor  = blue,
            citecolor = blue,
            anchorcolor = blue]{hyperref}\usepackage{ragged2e}
\usepackage{algorithm2e}
\usepackage[affil-it]{authblk}
\usepackage{abstract}

\usepackage[nodisplayskipstretch]{setspace}
\providecommand{\keywords}[1]
{
  \small	
  \textbf{\textit{Keywords---}} #1
}
\begin{document} 

\title{\Large{\textbf{
   Neutrino Mixing and Resonant Leptogenesis in Inverse Seesaw and  $\boldsymbol{\Delta(54)}$ Flavor Symmetry 
 \vspace{-0.45em}}}}

\author{Hrishi Bora$^1$%
\thanks{\href{mailto:hrishi@tezu.ernet.in}{hrishi@tezu.ernet.in} (Corresponding author)}}
  
   \author{Ng. K. Francis$^1$\thanks{\href{mailto:francis@tezu.ernet.in}{francis@tezu.ernet.in}}}
  \author{Bikash Thapa$^1$\thanks{\href{mailto:bikash2@tezu.ernet.in}{bikash2@tezu.ernet.in}}}
  \author{Shawan Kumar Jha$^2$\thanks{\href{mailto:francis@tezu.ernet.in}{shawankumar@iitg.ac.in}}}

\affil{\vspace{-1.05em} $^1$Department of Physics, Tezpur University, Tezpur - 784028, India}
\affil{\vspace{-1.05em}$^2$Department of Physics, Indian Institute of Technology Guwahati,
Guwahati 781039, India}

\date{\vspace{-5ex}}
\maketitle
\begin{center}
\textbf{\large{Abstract}}\\
\justify

The current work involves augmenting the $\Delta(54)$ discrete flavor model by incorporating two Standard Model Higgs particles into the Inverse Seesaw mechanism. We introduced Weyl fermions and Vector like fermions, which are gauge singlets in the Standard Model and produces Majorana mass terms in our lagrangian. The resulting mass matrix deviates from  the tribimaximal neutrino mixing pattern producing a non-zero reactor angle ($\theta_{13}$) . We have determined the effective Majorana neutrino mass, which is the parameter of relevance in neutrinoless double beta decay investigations, using the model's limited six-dimensional parameter space. We additionally investigate the possibility of baryogenesis in the proposed framework via resonant leptogenesis.  We have the non-zero value for resonantly enhanced CP asymmetry originating from the decay of right-handed neutrinos at the TeV scale, accounting for flavor effects. The evolution of lepton asymmetry is systematically analyzed by numerically solving a set of Boltzmann equations, leading to the determination of the baryon asymmetry with a magnitude of $ \lvert \eta_B \rvert \approx 6 \times 10^{-10}$. This outcome is achieved by selecting specific values for the right-handed neutrino mass $M_1 = 10$ TeV and mass splitting, $d \approx 10^{-8}$.
\end{center}


\hspace{-1.9em} PACS numbers: 12.60.-i, 14.60.Pq, 14.60.St
\newpage
\section{Introduction}
\label{sec:intro}

The neutrino masses along with their flavor mixing as observed in neutrino oscillations, leads to a question about where these tiny masses come from \cite{aker2019improved, faessler2020status, araki2005measurement, cao2021physics, nath2021detection}. Since the standard model do not include right-handed neutrinos unlike other fermions, it is unlikely that neutrino masses work the same way as the masses of charged fermions. The origin of neutrino masses can be explained by various frameworks beyond the standard model (BSM), including the  seesaw mechanism \cite{minkowski1977mu, king2005testing, mohapatra1986mechanism}, radiative seesaw mechanism \cite{ma2009radiative}, extra-dimensional models \cite{mohapatra1999neutrino, arkani2001neutrino } and others. The extension of the standard model with Inverse Seesaw mechanism can explain the observed Baryon Asymmetry of the Universe (BAU) through leptogenesis \cite{fukugita1986barygenesis}. These references includes many current reviews on neutrino physics\cite{Nguyen:2018rlb, King:2014nza,King:2003jb,Cao:2020ans, Ahn:2014zja, mcdonald2016nobel, Nguyen:2020ehj, Hong:2022xjg, kajita2016nobel,de20212020,okada2021spontaneous, Ahn:2021ndu, PhongNguyen:2017meq,Buravov:2014dna, buravov2009elementary, barman2023neutrino, bora2023majorana}.

In leptogenesis, the asymmetry in leptons, obtained from the CP-violating decay of heavy right-handed neutrinos, is transformed into an asymmetry in baryons through sphaleron processes \cite{kuzmin1985anomalous}. According to Ref. \cite{davidson2002lower},  a mass scale of around $\mathcal{O}(10^9)$ for the right-handed neutrino is necessary to explain the observed BAU. However, this requirement can be lessened if the masses of right-handed neutrinos are nearly the same. In such cases, the effects that violate CP symmetry become significantly amplified, and with relatively low masses (TeV scale), sufficient asymmetry in leptons can be generated to account for the Baryon Asymmetry of the Universe (BAU). This condition is termed resonant leptogenesis. It is important to mention that recent research, utilizing the $SU(5) \times \mathcal{T}_{13}$ model \cite{fong2021low}, has shown the possibility of resonant leptogenesis at the GeV–TeV scale within the type-I seesaw model, considering active sterile mixing within the sensitivity range of DUNE. Additionally, considerable attention has been devoted over time to investigating the origin of neutrino flavor mixing. Among the available explanations, Tri-bimaximal mixing (TBM) appears to be the most probable. However, experimental results from Daya Bay, RENO, and Double Chooz suggest that TBM needs to be adjusted to incorporate a non-zero value for $\theta_{13}$.
In this study, we introduce a model constructed within the minimum seesaw model framework, employing $\Delta(54)$ discrete symmetry. By employing resonant leptogenesis, the resulting mass matrix can potentially explain the Baryon Asymmetry of the Universe (BAU) concurrently. To achieve successful resonant leptogenesis, we introduce a higher-order term. We specifically choose the Majorana mass matrix for right-handed neutrinos, $M_{R}$, so that these neutrinos have degenerate masses at the dimension five-level. In essence, our work expands upon the model proposed in \cite{bora2024neutrino}, making it suitable for investigating resonant leptogenesis in scenarios involving the minimum seesaw model.

Similar study on resonant leptogenesis, utilizing $S_4$ symmetry within the minimum seesaw model has been carried out in Ref.\cite{thapa2021resonant}. However, in contrast to our current research, the models discussed in Ref.\cite{thapa2021resonant} achieve resonant leptogenesis differently. They achieve this by creating mass differences among the heavy right-handed neutrinos through minimal seesaw model based on $S_4$ discrete flavor symmetry that leads to TM$_{1}$ mixing. We explore the investigation of resonant leptogenesis within the inverse seesaw model based on $\Delta(54)$ discrete symmetry, while considering the discovery of a non-zero $\theta_{13}$.

This paper is structured as follows: In Section \ref{sec: frame}, we introduce the $\Delta(54)$ discrete symmetry with inverse seesaw mechanism and discuss the characteristics of the flavor group relevant to constructing the model. Section \ref{num} outlines the allowable range for model parameters based on the constraints imposed by the $3{\sigma}$ range of neutrino oscillation data. The numerical solution of the Boltzmann equations, which govern the evolution of lepton number density and the baryon asymmetry parameter, is presented in Section \ref{Res}, along with the framework for resonant leptogenesis. In Section \ref{conc}, we conclude our study and provide numerical results regarding neutrinoless double beta decay within the model.

\section{Framework of the Model}
\label{sec: frame}
Extending the fermion sector within the Standard Model framework is necessary to achieve the implementation of the Inverse Seesaw mechanism. Here, we have introduced Vector-like (VL) fermions, $N_1$ and $N_2$, which have the property of being gauge singlets inside the Standard Model framework. 
 After symmetry breaking, this new piece generates a Majorana mass ($M_N$) term which is negligible under the seesaw hierarchy  $M_N , M_S << m_{\nu N} << M_{NS} $. We also introduced a Weyl fermion denoted as $S_1$. In fact, the $\phi$
VEV induces a Majorana mass term for the $S_1$ fermion.  The fields associated with right-handedness and left-handedness are indicated by subscripts 1 and 2 respectively. The $\Delta(54)$  group includes irreducible representations  $1_1$, $1_2$, $2_1$, $2_2$, $2_3$, $2_4$, $3_{1(1)}$, $3_{1(2)}$, $3_{2(1)}$ and $3_{2(2)}$.  The products of  $3_{2(1)}\otimes3_{2(2)}$,  $3_{1(1)}\otimes3_{1(2)}$ , $3_{1(2)}\otimes3_{2(1)}$ and $3_{1(1)}\otimes3_{2(2)}$ lead to the trivial singlet.

The rules for multiplication are as follow \cite{ishimori2010non}:

$$3_{1(1)}\otimes3_{1(1)}=3_{1(2)} \oplus 3_{1(2)} \oplus3_{2(2)}$$
$$3_{1(2)}\otimes3_{1(2)}=3_{1(1)}\oplus 3_{1(1)}\oplus 3_{2(1)}$$
$$3_{2(1)}\otimes3_{2(1)}=3_{1(2)}\oplus 3_{1(2)}\oplus 3_{2(2)}$$
$$3_{2(2)}\otimes3_{2(2)}=3_{1(1)}\oplus 3_{1(1)}\oplus 3_{2(1)}$$
$$3_{1(1)}\otimes3_{1(2)}=1_{1}\oplus 2_{1}\oplus 2_{2}\oplus 2_{3}\oplus 2_{4}$$
$$3_{1(2)}\otimes3_{2(1)}=1_{2}\oplus 2_{1}\oplus 2_{2}\oplus 2_{3}\oplus 2_{4}$$
$$3_{2(1)}\otimes3_{2(2)}=1_{1}\oplus 2_{1}\oplus 2_{2}\oplus 2_{3}\oplus 2_{4}$$
$$3_{1(1)}\otimes3_{2(2)}=1_{2}\oplus 2_{1}\oplus 2_{2}\oplus 2_{3}\oplus 2_{4}$$

\begin{table}[ht]
    \centering
    \scalebox{0.8}{
 {\begin{tabular}{c c c c c  c c c c c c c c c c}
    \hline
       \textrm{Field}  &  L & $l $ & $H_1$ & $H_2$  & $N_{1}$ & $N_{2}$  & $S_1$ & $\chi$ & $\chi^\prime$ & $\zeta$ & $\zeta^\prime$ & $\xi$ & $\Phi_{S}$ & $\phi$ \\
     \hline
     \textrm{$\Delta(54)$}  &  $3_{1(1)}$ &  $3_{2(2)} $ & $1_{1}$ & $1_{2}$ & $3_{1(1)}$ & $3_{2(1)}$  & $3_{2(2)}$ & $1_2$ & $2_1$ & $1_{2}$ & $1_{1}$ & $3_{2(1)}$ & $3_{1(1)}$ & $3_{1(2)}$\\
     \textrm{Z}$_2$  &  1 & -1 & 1 & 1 & -1 & 1 & 1 & -1 & -1 & -1 & 1 & -1 & -1 & 1\\
    \textrm{Z}$_3$  &  $\omega$ & $\omega$ & 1 & 1  & 1 & $\omega$  & 1 & 1 & 1& 1 & $\omega$ & $\omega$ & $\omega$ & 1\\
\textrm{Z}$_4$  &  1 & -1 & 1 & 1  & 1 & -1  & 1 & -1 &-1& 1 & -1 & 1 & 1 & 1\\
\textrm{U(1)}  &  1 & 1 & 0 & 0  & 1 & 1  & 1 & 0 & 0 & 0 & 0 & 0 & 0& 0\\
     \hline
    \end{tabular}}}
    \caption{Particle content of our model}
    \label{tab:1}
    \end{table}

 We have developed a model based on the $\Delta(54)$ model, which incorporates the inclusion of additional flavons, namely $\chi$, $\chi^\prime$, $\xi$, $\zeta$, $\zeta^\prime$, $\Phi_{S}$, and $\phi$. In order to avoid undesired terms, we added additional symmetry $Z_2\otimes Z_3 \otimes Z_4$. Details on the particle composition and associated charge assignment according to the symmetry group are given in Table \ref{tab:1}. The left-handed leptons doublets and the right-handed charged lepton are assigned using the triplet representation of $\Delta(54)$.

The Lagrangian is as follows   :
 \begin{align}
  \mathcal{L} = & \frac{y_1}{\Lambda} ( l \Bar{L} ) \chi H_1 + \frac{y_2}{\Lambda} ( l \Bar{L} ) \chi^\prime H_1   + \frac{\Bar{L} \Tilde{H_2} N_{1}}{\Lambda}y_{\xi} \xi + \frac{\Bar{L} \Tilde{H_1} N_{1}}{\Lambda} y_{s} \Phi_{s} + \frac{\Bar{L} \Tilde{H_2} N_{1}}{\Lambda} y_{a} \Phi_{s}  \nonumber  \\                 
 & + y_{\scaleto{NS}{4pt}}  \Bar{N_{1}}S_{1}  \zeta + y_{\scaleto{NS}{4pt}} ^{\prime} \Bar{N_{2}}S_{1}  \zeta^\prime + \frac{y_{s_1}}{\Lambda}\Bar{S_{1}} S_{1} \phi + \frac{y_{s_2}}{\Lambda^2}\Bar{S_{1}} S_{1} \phi
 \label{eq1}
\end{align}

The vacuum expectation values are considered naturally as,
\begin{align*}
\langle \chi \rangle& =(v_{\chi})&
\langle \chi^\prime \rangle& =(v_{{\chi}^\prime},v_{{\chi}^\prime})&
\langle \xi \rangle& =(v_{\xi}, v_{\xi}, v_{\xi})&
\langle \phi \rangle& =(v_{\phi},v_{\phi},v_{\phi})&\\
\langle \Phi_S \rangle& =(v_s ,v_s,v_s)&
\langle \zeta \rangle& =(v_{\zeta})&
\langle \zeta^\prime \rangle& =(v_{\zeta}^\prime) &  
\end{align*}

The charged lepton mass matrix is given as \cite{ishimori2009lepton} 
\begin{align*}
    M_l= \frac{y_1 v}  {\Lambda}
    \begin{pmatrix}
    v_{\chi} & 0 & 0\\
    0 & v_{\chi}  & 0\\
    0 & 0 &  v_{\chi} 
    \end{pmatrix}  +  
    \frac{y_2 v}  {\Lambda}
    \begin{pmatrix}
    -\omega v_{\chi^\prime} + v_{\chi^\prime} & 0 & 0\\
    0 & -\omega^2 v_{\chi^\prime} + \omega^2 v_{\chi^\prime}  & 0\\
    0 & 0 &   -v_{\chi^\prime} + \omega v_{\chi^\prime}
    \end{pmatrix} 
\end{align*}

where $y_1$ and $y_2$ are coupling constants.

\subsection{Effective neutrino mass matrix}

After applying $\Delta(54)$ and electroweak symmetry breaking, the mass matrices related to the neutrino sector may be derived using the above-mentioned Lagrangian. The fundamental assumption of the ISS theory is the small $M_{S}$ scale, which guarantees small neutrino masses. The $M_{S}$ scale must be at the KeV level in order to decrease the right-handed neutrino masses to the TeV scale. The inverse seesaw model is a TeV-scale seesaw model that maintains compatibility with light neutrino masses in the sub-eV range while allowing heavy neutrinos to remain as light as a TeV and Dirac masses to be as large as those of charged leptons.

\begin{equation}
M_{NS}=   y_{\scaleto{NS}{3pt}} \begin{pmatrix}
     v_{\zeta} & 0 & 0\\
    0 &  v_{\zeta} & 0\\
    0 & 0 &  v_{\zeta}
    \end{pmatrix}
\end{equation}

\begin{equation}    
M_{S}=  \frac{y_{s_1}}{\Lambda^{2}}\begin{pmatrix}
     v_{\phi} & 0 & 0\\
    0 &  v_{\phi} & 0\\
    0 & 0 &   v_{\phi}
    \end{pmatrix} 
\end{equation}

\begin{equation}
    M_{NS}^{\prime}=  y_{\scaleto{NS}{3pt}}^{\prime}
 \begin{pmatrix}
    v^\prime_{\zeta} & 0 & 0\\
    0 & v^\prime_{\zeta} & 0\\
    0 & 0 &  v^\prime_{\zeta}
    \end{pmatrix}
    \end{equation}

 \begin{equation}
 M_{\nu N}= \frac{v}{\Lambda}        \begin{pmatrix} 
    y_{\scaleto{\xi}{6pt}} v_{\scaleto{\xi}{6pt}} & y_{s}v_{s} + y_{a} v_a & y_{s}v_{s} - y_{a} v_a\\
   y_{s}v_{s} - y_{a} v_a &   y_{\scaleto{\xi}{6pt}} v_{\scaleto{\xi}{6pt}} & y_{s}v_{s} + y_{a} v_a\\
    y_{s}v_{s} + y_{a} v_a & y_{s}v_{s} - y_{a} v_a &   y_{\scaleto{\xi}{6pt}} v_{\scaleto{\xi}{6pt}}
    \end{pmatrix}
   \end{equation}

In the inverse seesaw framework, the effective neutrino mass matrix can be written as
\begin{equation} 
m_\nu = M_{\nu N}(M_{NS}^{\prime})^{-1}M_{S}M_{NS}^{-1}M^\prime_{\nu N}
\label{eq6}
\end{equation}
\begin{equation} 
m_\nu = M_{\nu N} M^{-1}_{mid} M_{\nu N}^T
  \quad \text{with} \quad M_{mid} = M_{NS}^{\prime}M_{S}^{-1}M_{NS} 
\end{equation} 

\begin{equation}
     M_{mid}=
    \begin{pmatrix}
     M  &  0  &  0\\
     0  &  M  & 0  \\
     0 &   0   & M
    \end{pmatrix}
    \label{eq8}
\end{equation}

\begin{equation}
     M_{\nu N}=
    \begin{pmatrix}
     x  &  c+a  &  c-a\\
     c-a  &  x  & c+a  \\
     c+a &   c-a   & x
    \end{pmatrix} 
    \label{eq9}
\end{equation}

where
$x=y_{\scaleto{\xi}{6pt}} v_{\scaleto{\xi}{6pt}}$,
$c =y_s v_s$, $a=y_a v_s$, $M =  \frac{\Lambda^2 y_{\scaleto{NS}{3pt}} y{^\prime}_{\scaleto{NS}{3pt}} v_\zeta    v^\prime_\zeta}{v_\phi y_{s_1}}$.

The resultant mass matrix from Eq. (\ref{eq6})
\begin{equation}
    m_\nu= \frac{v^2}{M \Lambda^2}
    \begin{pmatrix}
     2a^2 + 2c^2 + x^2   &   -a^2 + c^2 + 2cx   &      -a^2 + c^2 + 2cx\\
     -a^2 + c^2 + 2cx    &  2a^2 + 2c^2 + x^2   &       -a^2 + c^2 + 2cx \\
    -a^2 + c^2 + 2cx &   -a^2 + c^2 + 2cx   &    2a^2 + 2c^2 + x^2 
    \end{pmatrix}
\end{equation}

\begin{equation}
\label{eq10}
    m_\nu= \frac{1}{M}
    \begin{pmatrix}
     2a^{\prime2} + 2c^{\prime2} + x^{\prime2}   &   -a^{\prime2} + c^{\prime2} + 2c^{\prime} x^{\prime}   &     -a^{\prime2} + c^{\prime2} + 2c^{\prime} x^{\prime}\\
     -a^{\prime2} + c^{\prime2} + 2c^{\prime} x^{\prime}    &  2a^{\prime2} + 2c^{\prime2} + x^{\prime2}   &       -a^{\prime2} + c^{\prime2} + 2c^{\prime} x^{\prime} \\
   -a^{\prime2} + c^{\prime2} + 2c^{\prime} x^{\prime} &   -a^{\prime2} + c^{\prime2} + 2c^{\prime} x^{\prime}  &    2a^{\prime2} + 2c^{\prime2} + x^{\prime2}
    \end{pmatrix}
\end{equation}

where  $a^{\prime} = \frac{av}{\Lambda}$,  $c^{\prime} = \frac{cv}{\Lambda}$,  $x^{\prime} = \frac{xv}{\Lambda}$. The dimension of the problem is absorbed by the term $M$ and the components of the matrix are unaffected . Phase redefinitions of the charged lepton fields can absorb the phase of $M$, allowing it to be considered as a real parameter without losing generality \cite{lei2020minimally}.\\
In the following sections, we have presented the numerical
approaches and discussed baryogenesis via resonant leptogenesis, within the context of
our model.

\section{Numerical Analysis} 
\label{num}

\noindent The
mass matrix in Eq. (\ref{eq10}) gives the effective neutrino mass
matrix in terms of the model complex parameters $a^\prime$, $c^\prime$, and
$x^\prime$. We find the values of the model parameters by fitting the
model to the current neutrino oscillation data.  We
use the 3$\sigma$ interval for the neutrino oscillation parameters
( $\theta_{12}, \theta_{23}, \theta_{13}, \Delta m^2_{21}
, \Delta m^2_{31}$ ) as presented in Table \ref{tab:2}. A
further constraint on the model parameters was applied on
the sum of absolute neutrino masses	from the cosmological bound $\sum_{i} m_i < 0.12 eV$. In our analysis, the three complex parameters of the model
are treated as free parameters and are allowed to run over the
following ranges:
$\lvert a^\prime \rvert \in [0, 0.3] eV $, $ \phi_a \in [-\pi , \pi]$ ;
$\lvert c^\prime \rvert \in [0, 10^{-1}] eV$ ; $\phi_c \in [-\pi , \pi]$ ;
$\lvert x^\prime \rvert \in [0, 10^{-3}] eV$ , $\phi_x \in [-\pi , \pi]$\\
where $\phi_a$ , $\phi_c$ and $\phi_x$ are the phases.

\begin{table}[ht]
\centering
 \scalebox{1}{
  \begin{tabular}{ | l | c | r |}
    \hline
    Parameters & NH (3$\sigma$) & IH (3$\sigma$) \\ \hline
    $\Delta{m}^{2}_{21}[{10}^{-5}eV^{2}]$ & $6.82 \rightarrow 8.03$ & $6.82 \rightarrow 8.03$ \\ \hline
    $\Delta{m}^{2}_{31}[{10}^{-3}eV^{2}]$ & $2.428 \rightarrow 2.597$ & $-2.581 \rightarrow -2.408 $\\ \hline
    $\sin^{2}\theta_{12}$ & $0.270 \rightarrow 0.341$ & $0.270 \rightarrow 0.341$ \\ \hline
     $\sin^{2}\theta_{13}$ & $0.02029 \rightarrow 0.02391$ & $0.02047 \rightarrow 0.02396$ \\ \hline
    $\sin^{2}\theta_{23}$ & $0.406 \rightarrow 0.620$ & $0.410 \rightarrow 0.623$ \\ \hline
    $\delta_{CP}$ & $108 \rightarrow 404$ & $192 \rightarrow 360$ \\ \hline
  \end{tabular}}
  \caption{ The neutrino oscillation parameters from NuFIT 5.2 (2022)}
    \label{tab:2}
\end{table}

\begin{figure}[ht]
     \centering
     \begin{subfigure}[b]{0.46\textwidth}
         \centering
         \includegraphics[width=\textwidth]{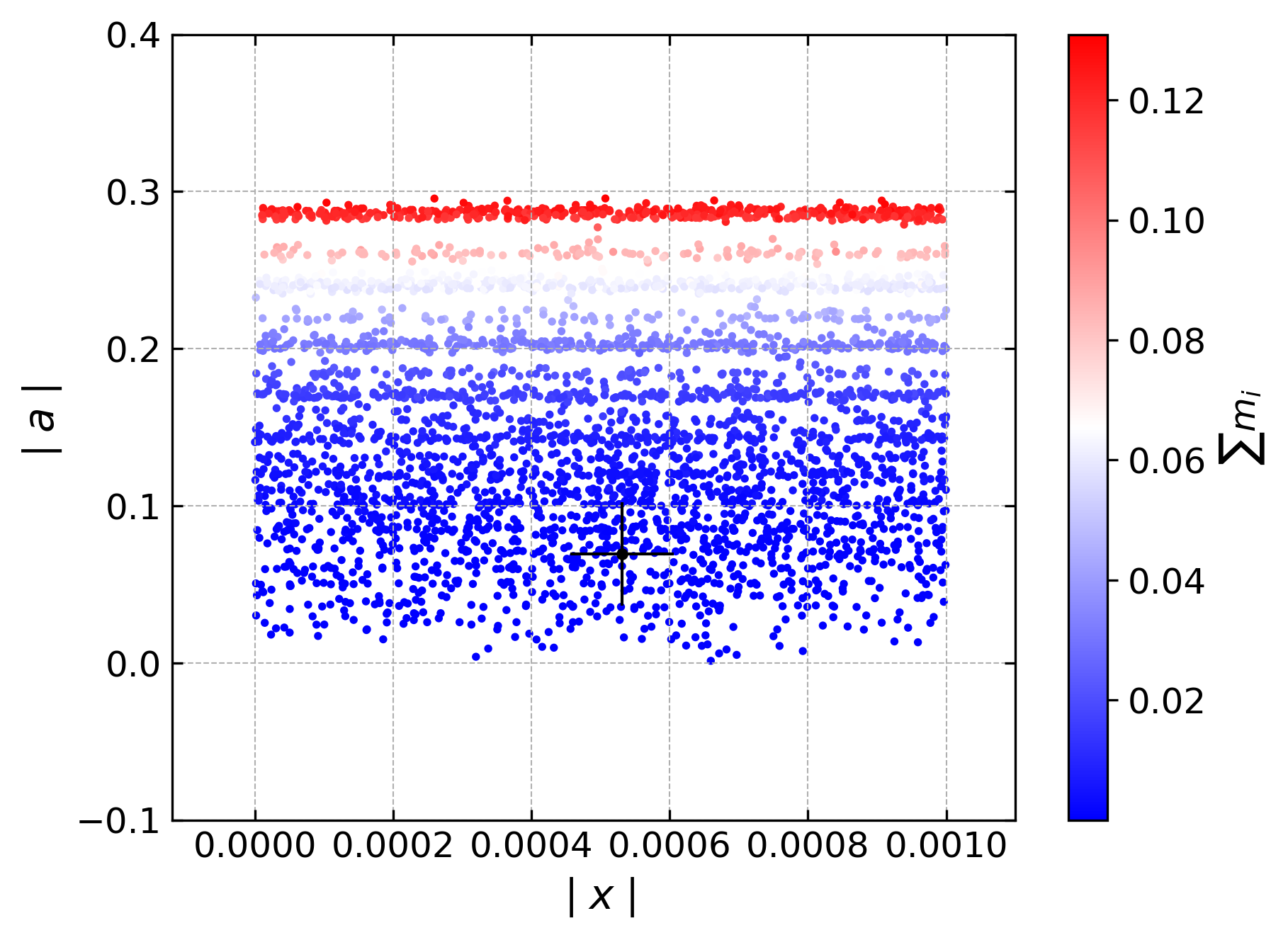}
     \end{subfigure}
     \hfill
     \begin{subfigure}[b]{0.46\textwidth}
         \centering
         \includegraphics[width=\textwidth]{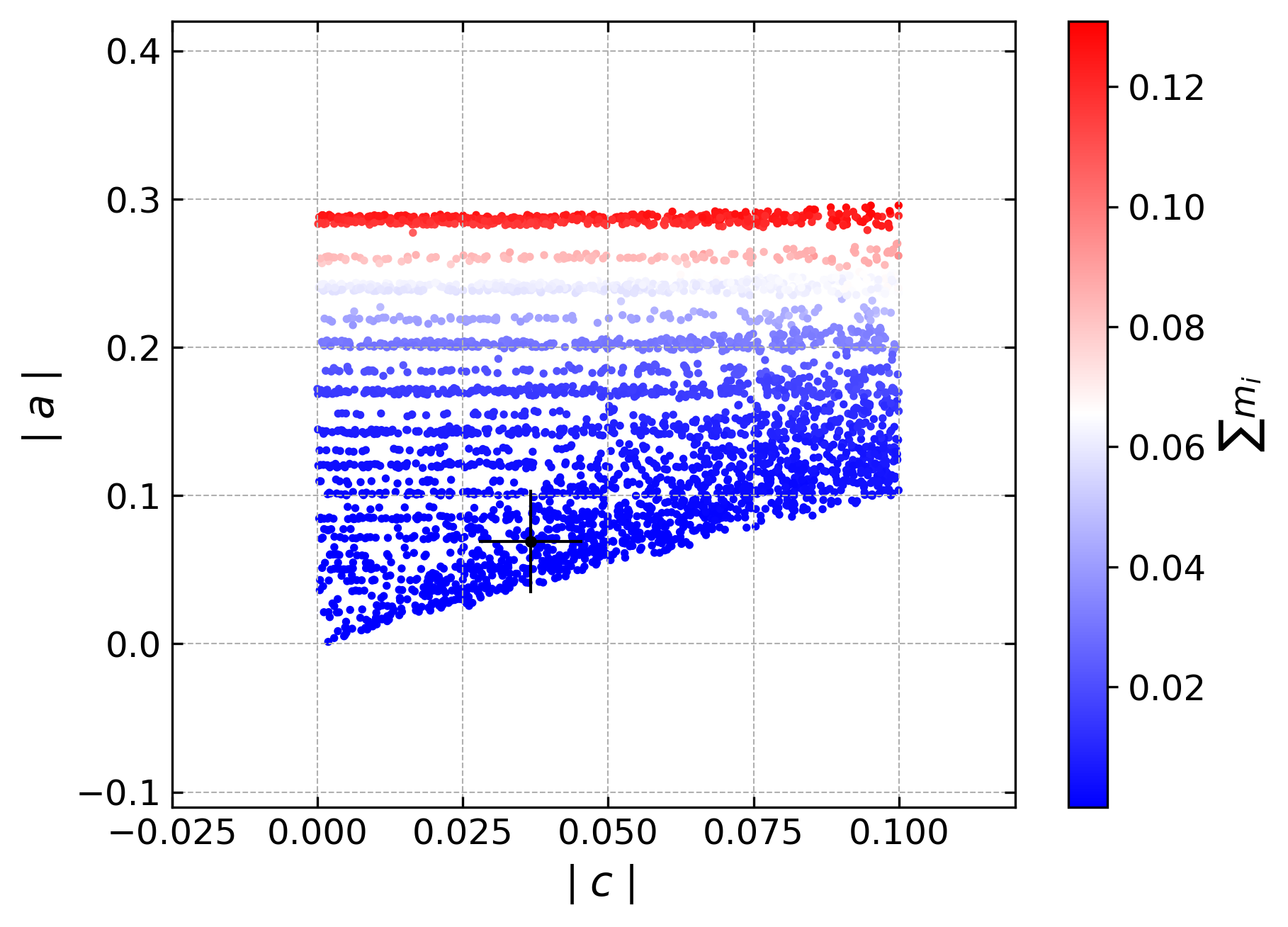}
     \end{subfigure}
      \caption{Left and right panel shows the correlation of model parameters along with the variation of $\sum m_i $. The black marker indicate
the best-fit values.}
    \label{fig:1}
\end{figure}

\begin{figure}[ht]
     \centering
     \begin{subfigure}[b]{0.46\textwidth}
         \centering
         \includegraphics[width=\textwidth]{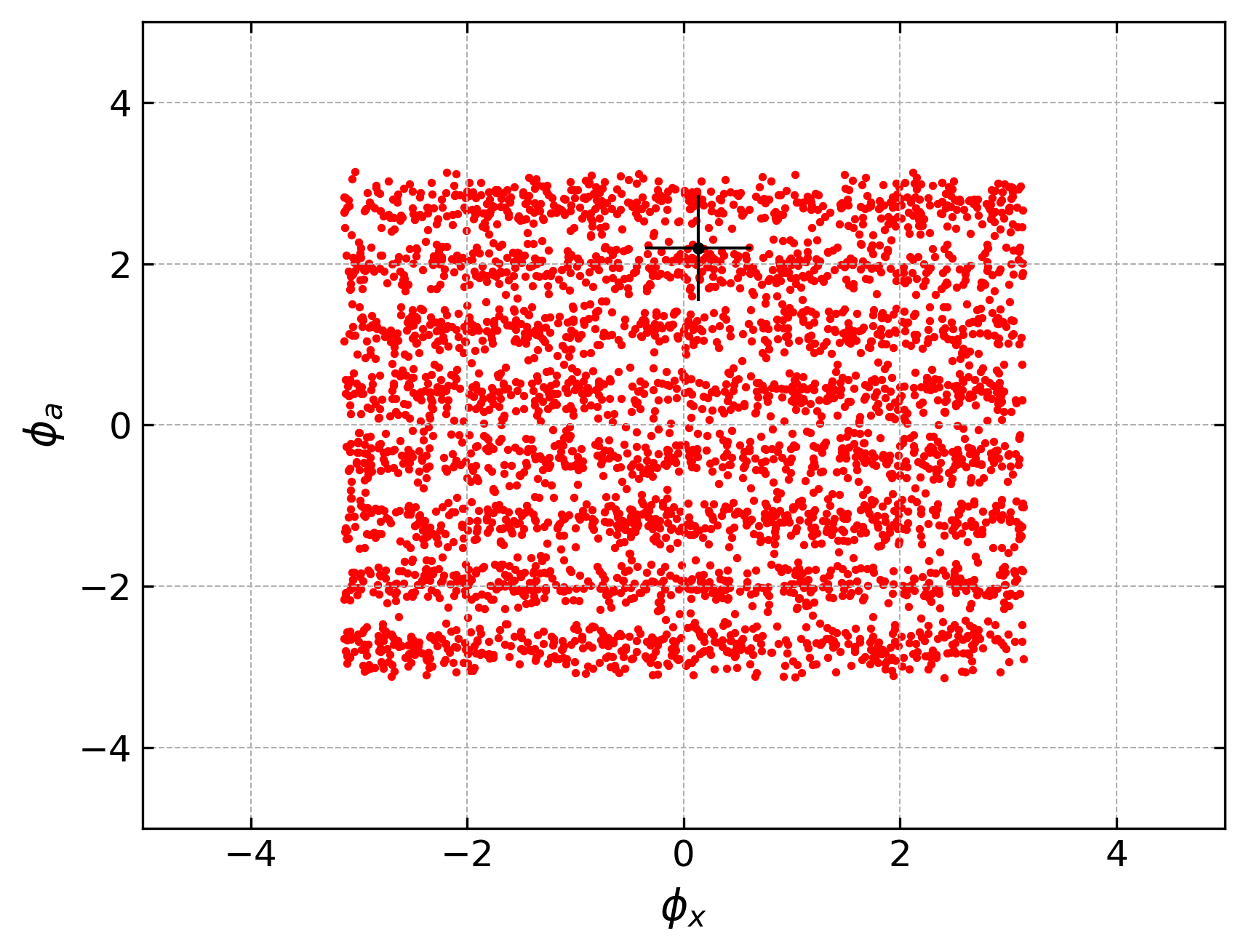}
     \end{subfigure}
     \hfill
     \begin{subfigure}[b]{0.46\textwidth}
         \centering
         \includegraphics[width=\textwidth]{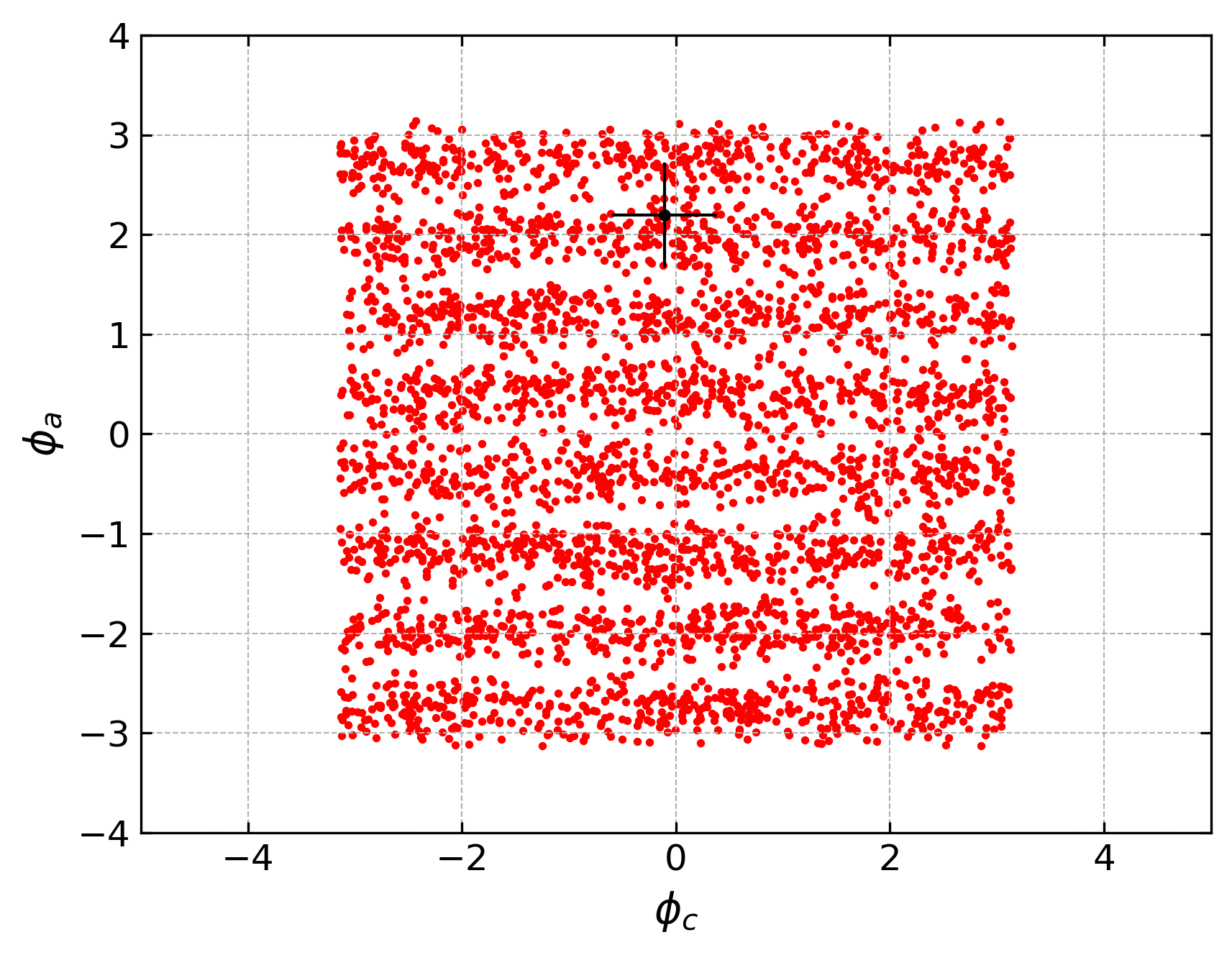}
     \end{subfigure}
      \caption{Correlation between the phases $\phi_x$ , $\phi_c$ and $\phi_a$. The black marker indicate the best-fit values corresponding to $\chi^2$ min.}
    \label{fig:2}
\end{figure}

The neutrino mass matrix $m_\nu$ is diagonalized by the PMNS matrix $U$ as follow \cite{lei2020minimally}:
\begin{equation}
    \label{eq:12}
    U^\dagger m^{(i)}_\nu U^* = \textrm{diag(}m_1, m_2, m_3 \textrm{)}
\end{equation}

 We numerically calculated $U$ using the relation $U^\dagger M_\nu U = \textrm{diag(}m_1^2, m_2^2, m_3^2 \textrm{)}$, where $M_\nu = m_\nu m^{\dagger}_\nu$. The neutrino oscillation parameters $\theta_{12}$, $\theta_{13}$, $\theta_{23}$ and $\delta$ can be obtained from $U$ as 
\begin{equation}
    \label{eq:13}
    s_{12}^2 = \frac{\lvert U_{12}\rvert ^2}{1 - \lvert U_{13}\rvert ^2}, ~~~~~~ s_{13}^2 = \lvert U_{13}\rvert ^2, ~~~~~~ s_{23}^2 = \frac{\lvert U_{23}\rvert ^2}{1 - \lvert U_{13}\rvert ^2}
\end{equation}

and $\delta$ may be given by
\begin{equation}
    \label{eq:14}
    \delta = \textrm{sin}^{-1}\left(\frac{8 \, \textrm{Im(}h_{12}h_{23}h_{31}\textrm{)}}{P}\right)
\end{equation}
with 
\begin{equation}
    \label{eq:15}
     P = (m_2^2-m_1^2)(m_3^2-m_2^2)(m_3^2-m_1^2)\sin 2\theta_{12} \sin 2\theta_{23} \sin 2\theta_{13} \cos \theta_{13}
\end{equation}

We adjusted the modified $\Delta(54)$ model to suit the experimental data by minimizing the ensuing $\chi^2$ function in order to evaluate how the neutrino mixing parameters contrast with the most current experimental data:

\begin{equation}
	\label{eq:16}
	\chi^2 = \sum_{i}\left(\frac{\lambda_i^{model} - \lambda_i^{expt}}{\Delta \lambda_i}\right)^2,
\end{equation}

where $\lambda_i^{model}$ is the $i^{th}$ observable predicted by the model, $\lambda_i^{expt}$ stands for  $i^{th}$ experimental best-fit value and $\Delta \lambda_i$ is the 1$\sigma$ range of the observable.

The best-fit values for $\lvert x^\prime \rvert$, $\lvert c^\prime \rvert$, $\lvert a^\prime \rvert$, $\phi_x$, $\phi_c$ and $\phi_a$ obtained are (0.00054, 0.03667, 0.06914, 0.13258$\pi$, -0.10508$\pi$, 2.23488$\pi$).

Correspondingly, the best-fit values for
the neutrino oscillation parameters are: $sin^2\theta_{12} = 0.31940$, $\sin^2 \theta_{13} = 0.02394$, $\sin^2 \theta_{23} =  0.51231 $, $\sin \delta_{CP} = 0.094$. The best-fit values for other parameters, such as  $\Delta m^2_{21}/\Delta m^2_{31}$ is 0.029, are correspond to the $\chi^2$-minimum.

\begin{figure}[h]
     \centering
     \begin{subfigure}[b]{0.46\textwidth}
         \centering
         \includegraphics[width=\textwidth]{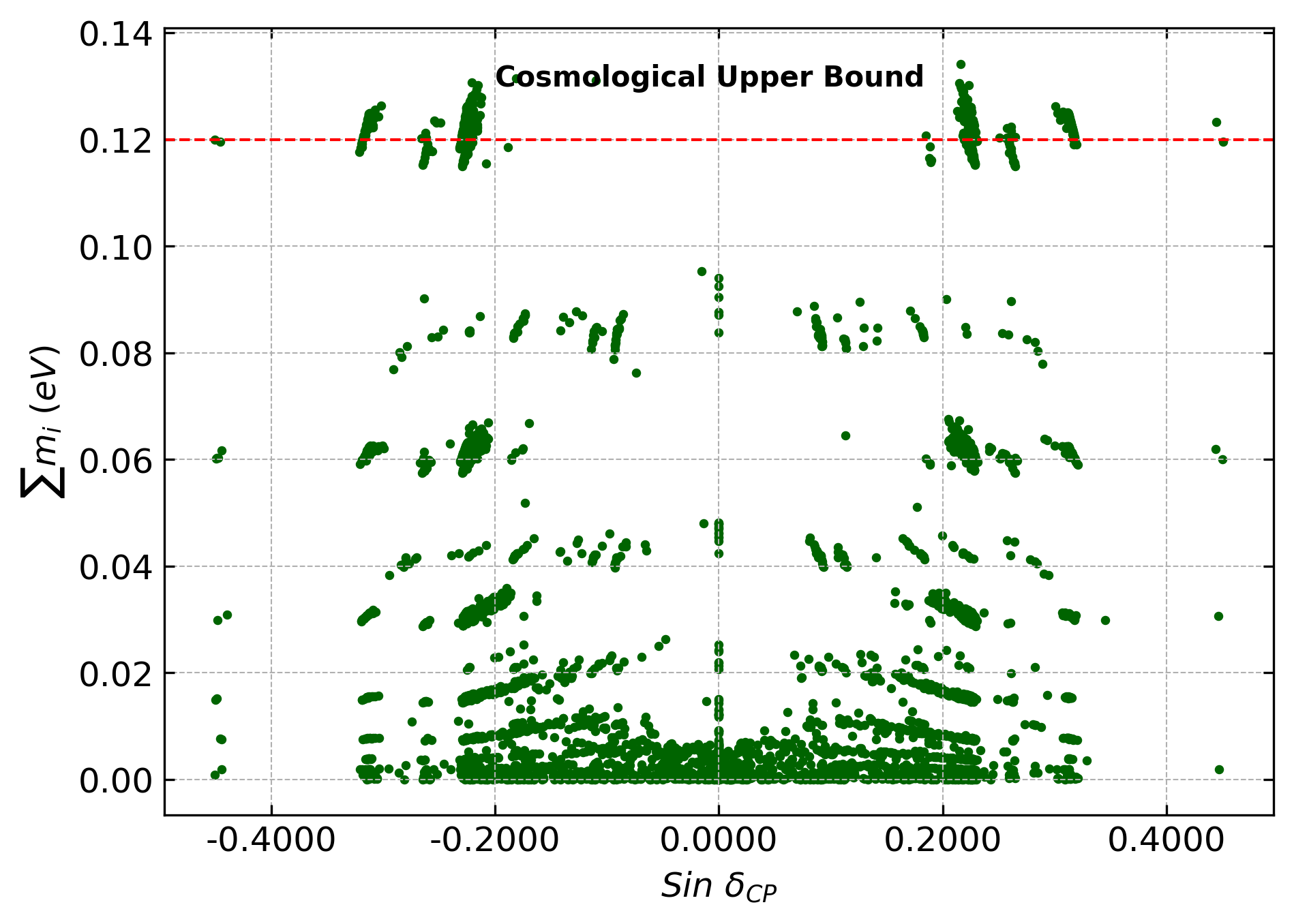}
     \end{subfigure}
     \caption{ Correlation between	$\sum m_i$ with sin $\delta_{CP}$. The red dotted horizontal line gives the upper limit of absolute neutrino mass.}
\end{figure}

\section{Resonant Leptogenesis} 
\label{Res}

Fukugita and Yanagida originally proposed the leptogenesis mechanism, which is one of the most commonly accepted explanations for the Baryon Asymmetry of the Universe(BAU). The mass of the lightest right-handed neutrino, $M_1 = 10^9 GeV$, has a lower bound in the simplest case of thermal leptogenesis with a hierarchical mass spectrum of right-handed neutrinos \cite{davidson2002lower}. Although one can lower this limit if their masses are nearly degenerate. This scenario is
popularly known as resonant leptogenesis\cite{pilaftsis2004resonant, pilaftsis1997cp}. In this scenario, the resonant enhancement amplifies the one-loop self-energy contribution, leading to the flavor-dependent asymmetry resulting from the decay of a right-handed neutrino into a lepton and Higgs.
\begin{equation}
\epsilon_{i\alpha} = \frac{{\Gamma(N_i \to l_\alpha + H) - \Gamma(N_i \to \bar{l}_\alpha + \bar{H})}}{{\sum_{\alpha} \left(\Gamma(N_i \to l_{\alpha} + H) + \Gamma(N_i \to \bar{l}_{\alpha} + \bar{H})\right)}} 
\end{equation}

\begin{equation}
\label{eq:3}
= \sum_{i \neq j}\frac{{  \text{Im}\biggl\{(Y^*_{\nu})_{ \alpha i} (Y_{\nu})_{\alpha j} \bigr[(Y_{\nu}^{\dagger} Y_{\nu})_{ij} + \xi_{ij} (Y_{\nu}^{\dagger} Y_{\nu})_{ji}}\bigr]\biggl\}} {(Y_{\nu}^{\dagger} Y_{\nu})_{ii} (Y_{\nu}^{\dagger} Y_{\nu})_{jj} } \times \frac{\xi_{ij} \zeta_j (\xi_{ij}^2 - 1)}{ (\xi_{ij} \zeta_j)^2 + (\xi_{ij}^2 - 1)^2}  
\end{equation}\\
where $\xi_{ij} = M_i/M_j$ and we took  $M_1 = 10 \ TeV $ and $d = (M_3 - M_1)/M_1 = 10^{-8}$

In our model, we have three right-handed neutrinos with exactly degenerate masses, $M_1 = M_2 = M_3 = M$. However, a tinymass separation between any two right-handed neutrinos is necessary for successful leptogenesis, and this is included to our model by including a higher dimension term in Eq.(\ref{eq1}). Such term leads to a minor correction in the Majorana mass matrix of Eq.(\ref{eq8}), and the resultant structure of the mass matrix may be written as

\begin{equation}
     M_{mid}=
    \begin{pmatrix}
     M  &  e  &  e\\
     e  &  M  & e  \\
     e &   e   & M
    \end{pmatrix} 
    \label{eq19}
\end{equation}

where $  e = \frac{y_{s_2}v_\phi}{\Lambda^2} $ is a parameter that quantifies the tiny
difference between masses required for leptogenesis. The
mass matrix in Eq.(\ref{eq19}) is diagonalized using a (3×3) matrix
of the form

\begin{equation*}
  D = \begin{pmatrix}
    -1  &  -1  &    1\\
     1  &  0 & 1  \\
    0 &   1   & 1 
    \end{pmatrix} 
\end{equation*}

with real eigenvalues $M_1 = M - e$ and $M_2 = M - e$ and $M_3 = M - 2e$         . In the
basis where the charged-lepton and Majorana mass matrix
are diagonal, the dirac mass matrix Eq.(\ref{eq9}) takes the form

\begin{equation*}
M^\prime_{\nu N} = M_{\nu N}.D = 
    \begin{pmatrix}
    \frac{v(a+s-x)}{\Lambda}  &  -\frac{v(a-s+x)}{\Lambda} &    \frac{v(2s+x)}{\Lambda}\\
     \frac{v(a-s+x)}{\Lambda}  &  \frac{v(a-v+x)}{\Lambda} & \frac{v(2s+x)}{\Lambda} \\
      -\frac{v(2a)}{\Lambda} &   -\frac{v(a+s-x)}{\Lambda}   & \frac{v(2s+x)}{\Lambda} 
    \end{pmatrix}  
\end{equation*}

From this point onward, we will take $Y_{\nu N} = M^\prime_{\nu N} / v $, which is
relevant for calculating CP asymmetry that arises during the
decay of right-handed neutrinos in out-of-equilibrium way.

\begin{figure}[ht]
     \centering
     \begin{subfigure}[b]{0.46\textwidth}
         \centering
         \includegraphics[width=\textwidth]{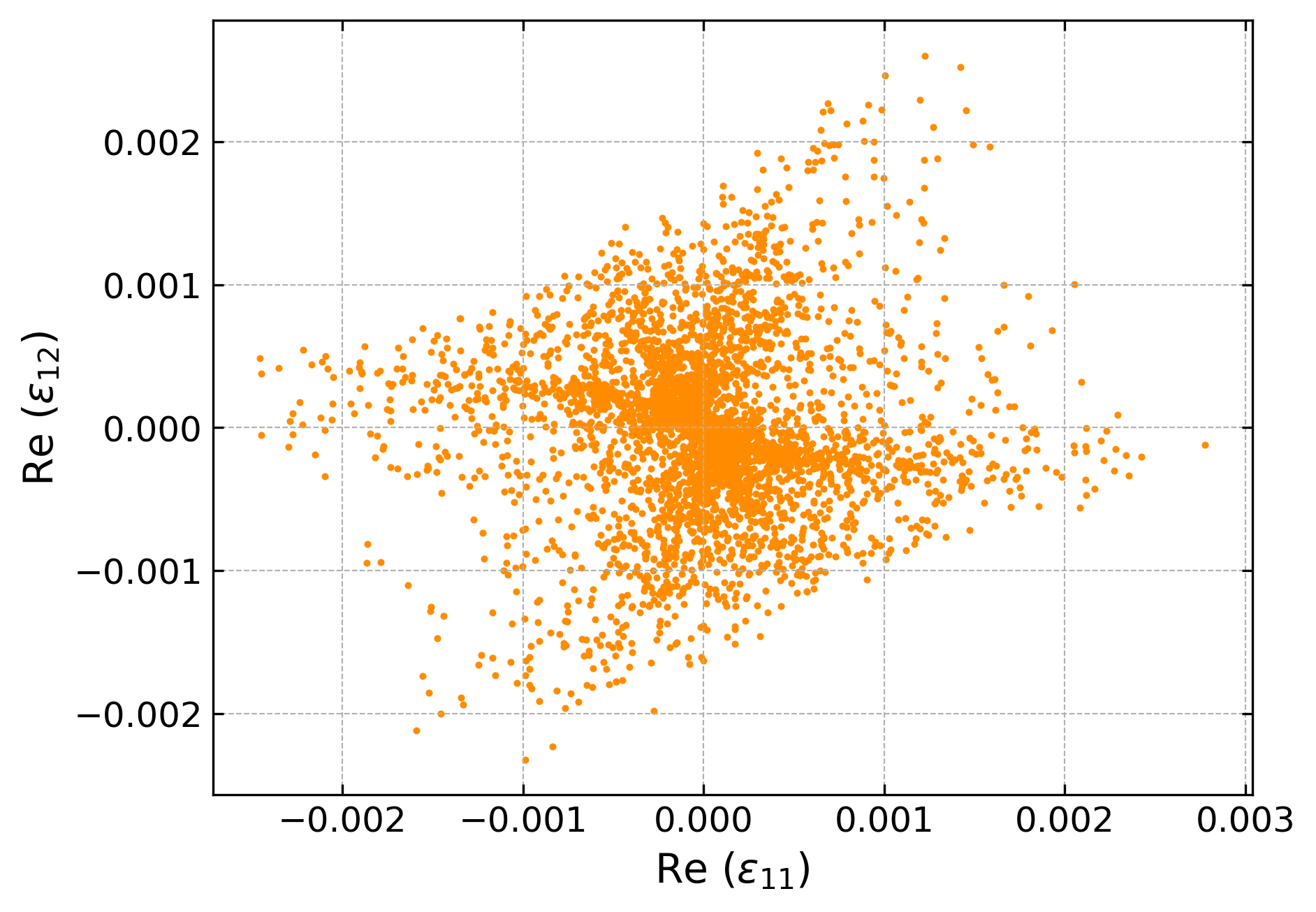}
     \end{subfigure}
     \hfill
     \begin{subfigure}[b]{0.46\textwidth}
         \centering
         \includegraphics[width=\textwidth]{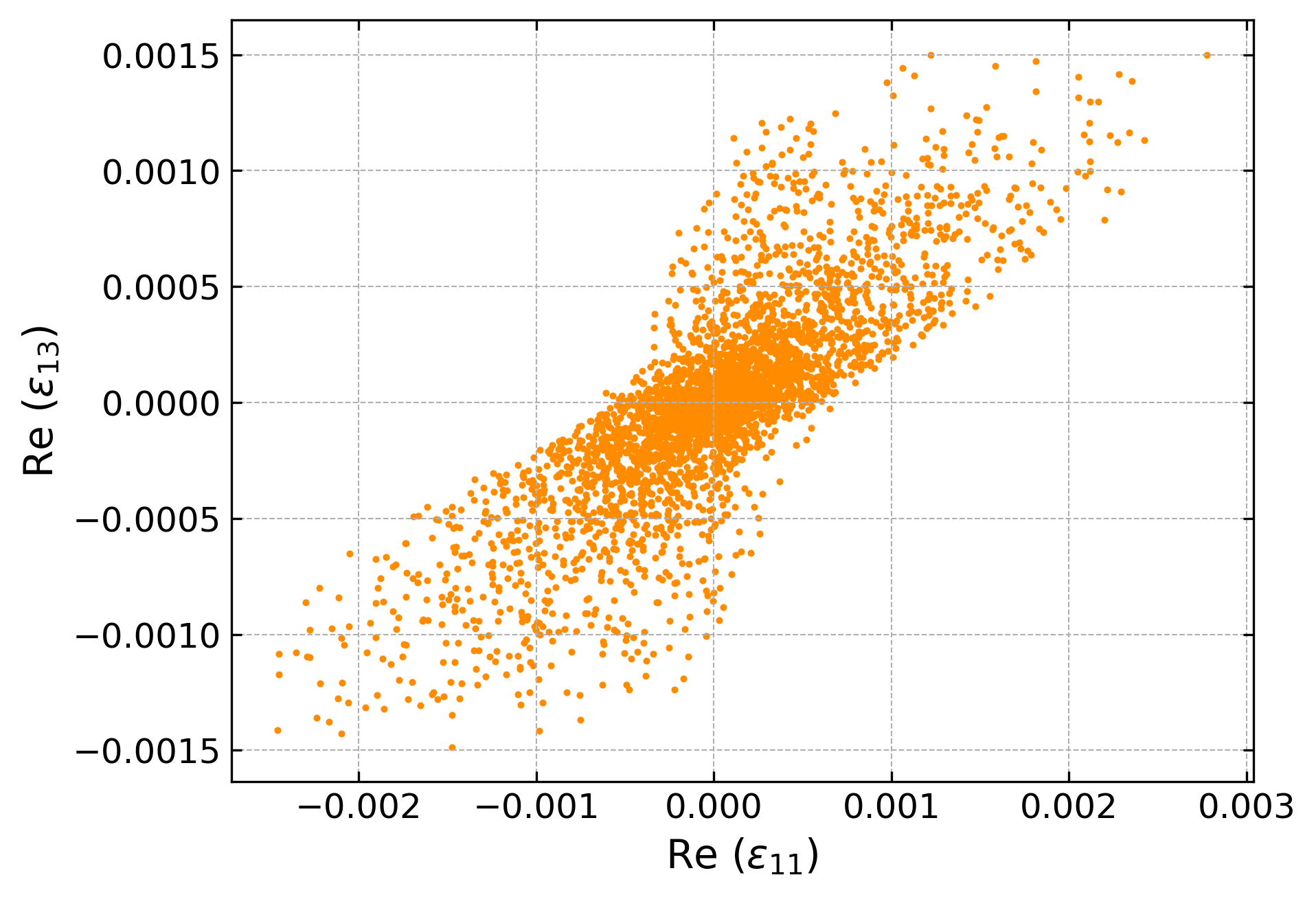}
     \end{subfigure}

      \begin{subfigure}[b]{0.46\textwidth}
         \centering
         \includegraphics[width=\textwidth]{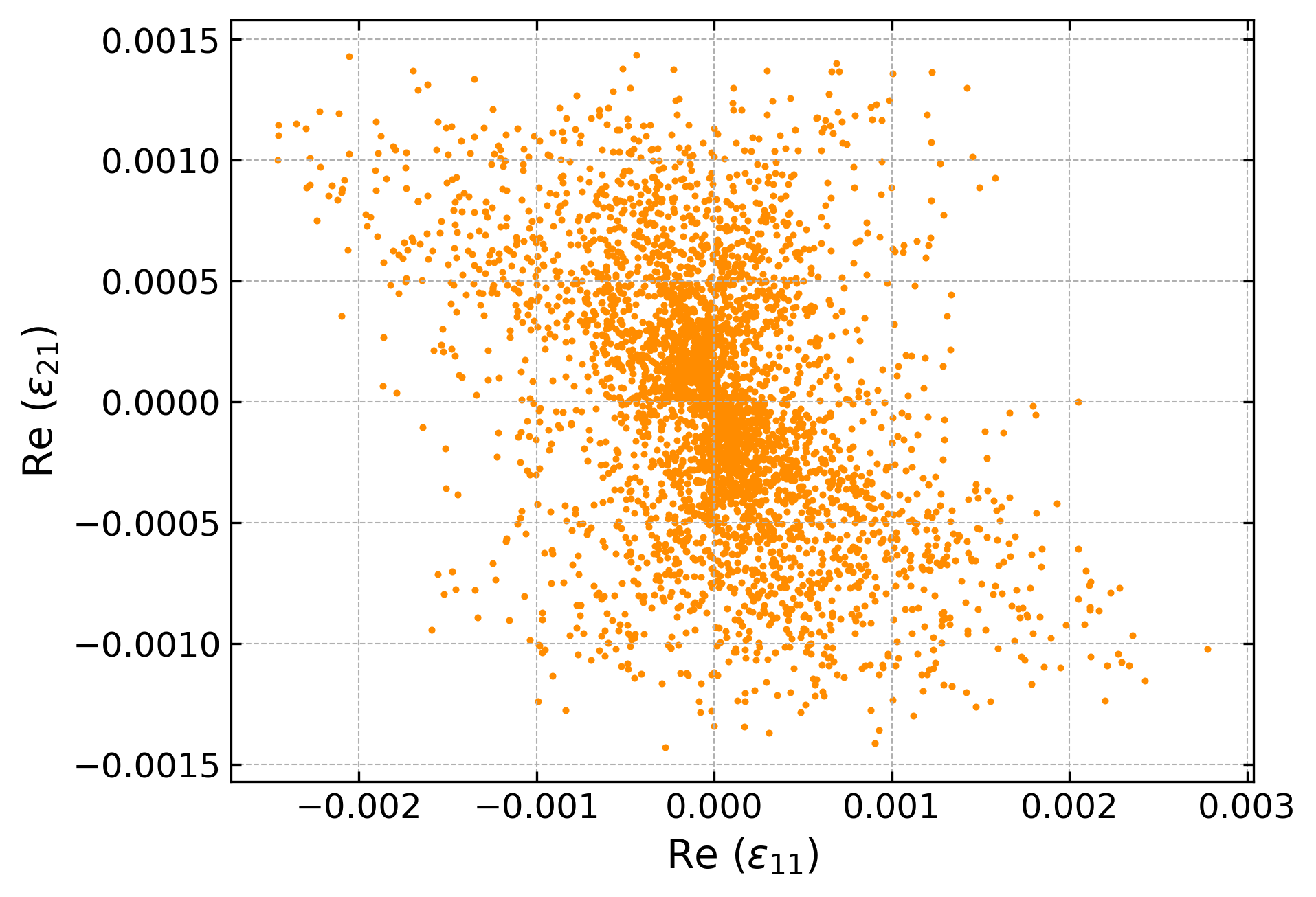}
     \end{subfigure}
     \hfill
     \begin{subfigure}[b]{0.46\textwidth}
         \centering
         \includegraphics[width=\textwidth]{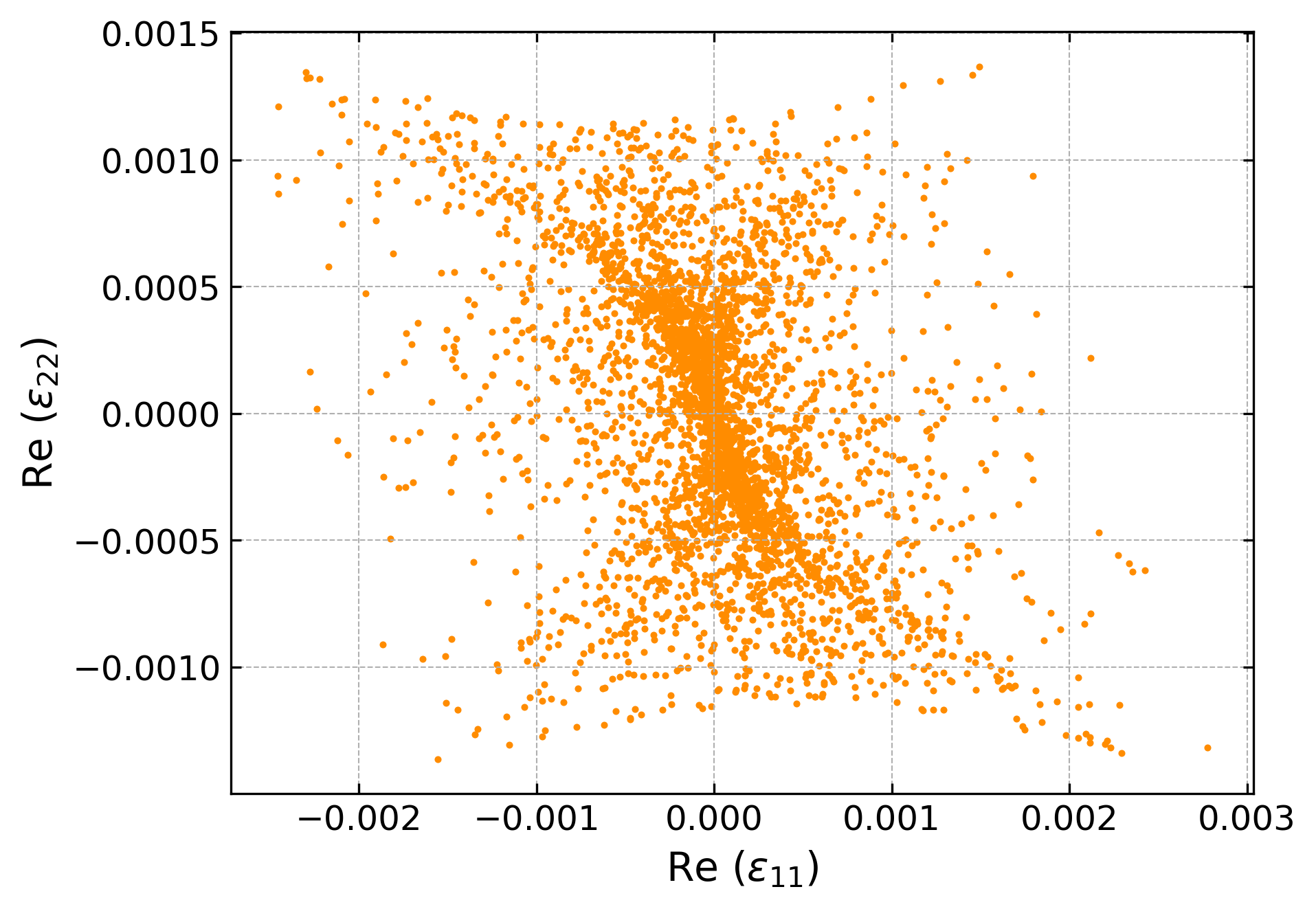}
     \end{subfigure}
  \caption{Correlations between the flavor-dependent CP-violating asymmetries $\varepsilon_{11}$ with $\varepsilon_{12}$, $\varepsilon_{21}$ and $\varepsilon_{22}$ respectively.}
  \label{fig4}
  \end{figure}

  The CP-violating asymmetries $\varepsilon_{i\alpha}$ in the flavored resonant leptogenesis scenario under study are related to the baryon-to-photon ratio $\eta_B$ as follows \cite{xing2020bridging}:

  \begin{equation}
      \eta_{B} \simeq -9.6 \times 10^{-3} \sum_{\alpha}(\varepsilon_{1\alpha}K_{1\alpha} + \varepsilon_{2\alpha}K_{2\alpha})
  \end{equation}
     
where $K_{1\alpha}$ and $K_{2\alpha}$ are the conversion efficiency factors. The region in which the lepton flavor takes effect determines the sum over the flavor index $\alpha$.
. To evaluate the sizes of $K_{i\alpha}$, let us first of all figure out the effective light neutrino masses.

\begin{equation}
    m_{i\alpha} \simeq \frac{v^2 \big\lvert (Y_v)_{\alpha i} \big\rvert^2}{M_i}
    \end{equation}

     The decay parameters $K_{i\alpha} \equiv \Tilde{m}_{i\alpha}/m_*$ can be calculated, where $m_* = 8\pi v^2 H(M_1)/M^2_1 \simeq 1.08 \times 10^{-3}eV$ gives the equilibrium neutrino mass and $H(M_1)$ is called the Hubble expansion parameter of the Universe.

    Now we define a dimensionless parameter $d \equiv (M_2 - M_1) / M_1 = \xi_{21} - 1$ to calculate the level of degeneracy for two of the three heavy Majorana neutrinos. Allowing for $d << 1$, we have $\kappa_{1\alpha} \simeq \kappa_{2\alpha} \equiv \kappa(K_\alpha)$ with $K_{\alpha} \equiv K_{1\alpha} + K_{2\alpha}$. Given the initial thermal abundance of heavy Majorana neutrinos, the efficiency factor $\kappa(K_\alpha)$ can be expressed as:
\begin{equation}
   \kappa(K_\alpha) \simeq \frac{2}{K_\alpha z_B(K_\alpha)} \biggr[ 1 - exp \biggl( \frac{-1}{2} K_\alpha z_B(K_\alpha) \biggl) \biggr]  
\end{equation}
where $z_B(K_\alpha) \simeq 2 + 4 K^{0.13}_\alpha exp(-2.5 / K_\alpha)$.

\begin{figure}[ht]
     \centering
     \begin{subfigure}[b]{0.46\textwidth}
         \centering
    \includegraphics[width=\textwidth]{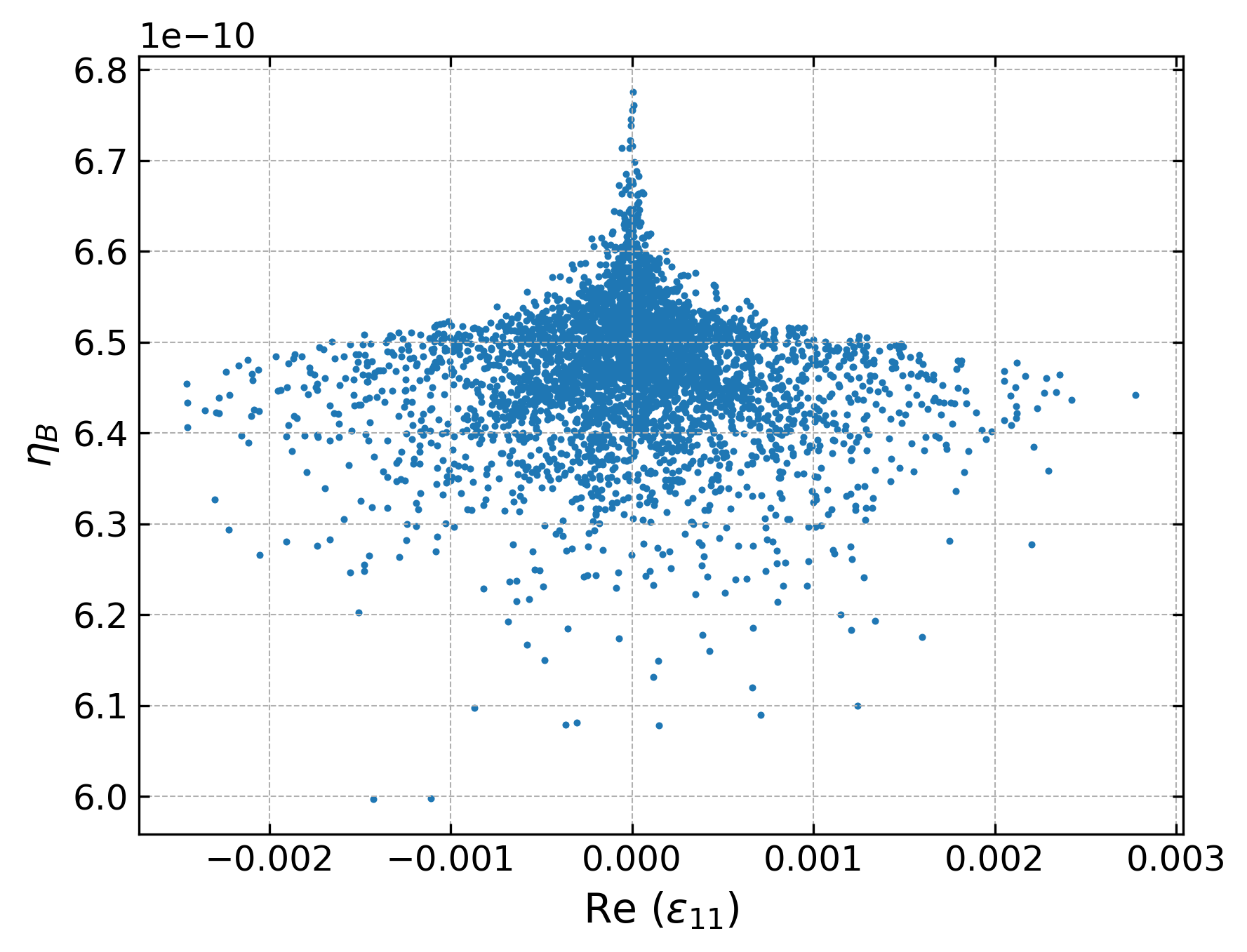}
     \end{subfigure}
     \hfill
     \begin{subfigure}[b]{0.46\textwidth}
         \centering
    \includegraphics[width=\textwidth]{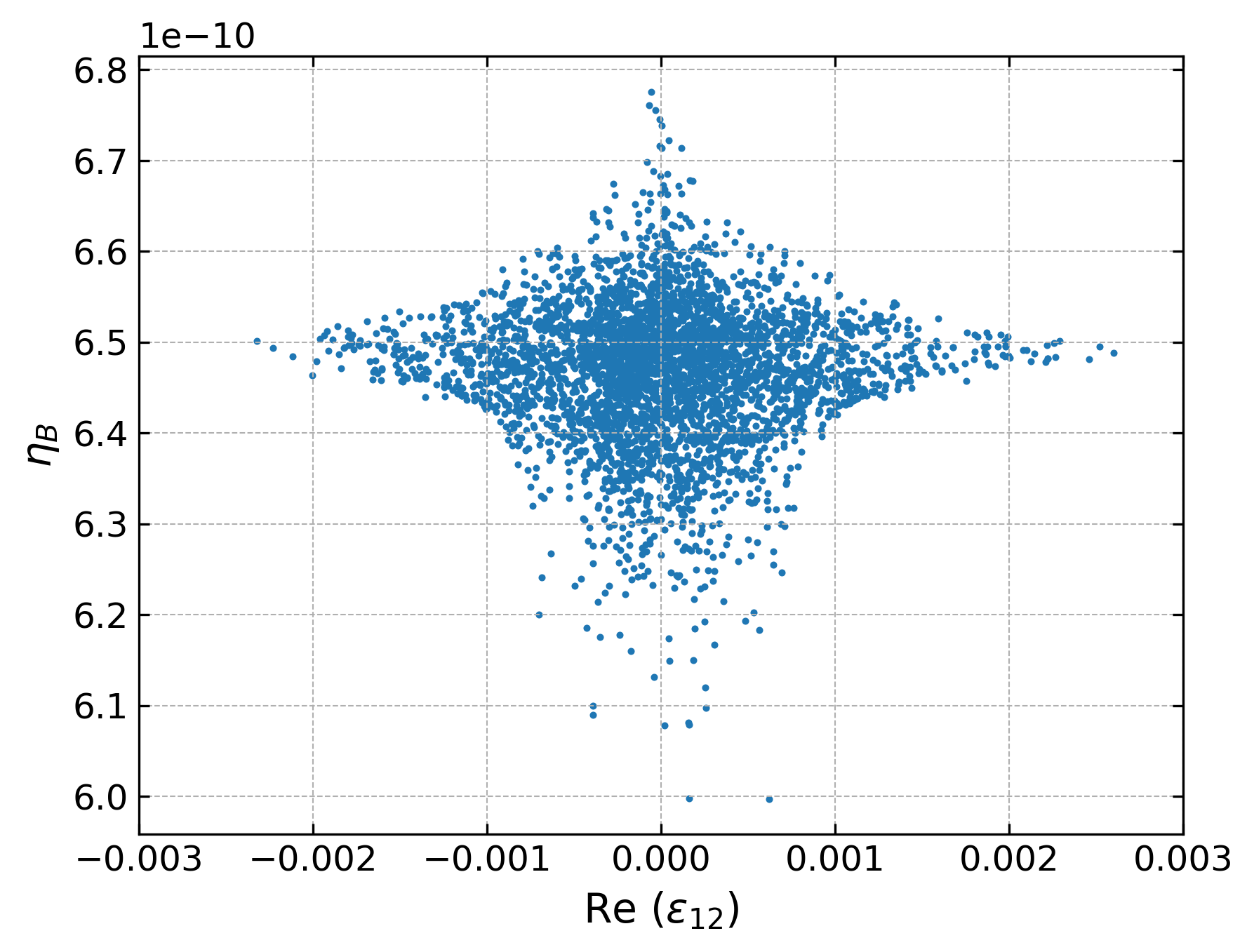}
     \end{subfigure}
 \begin{subfigure}[b]{0.46\textwidth}
         \centering
    \includegraphics[width=\textwidth]{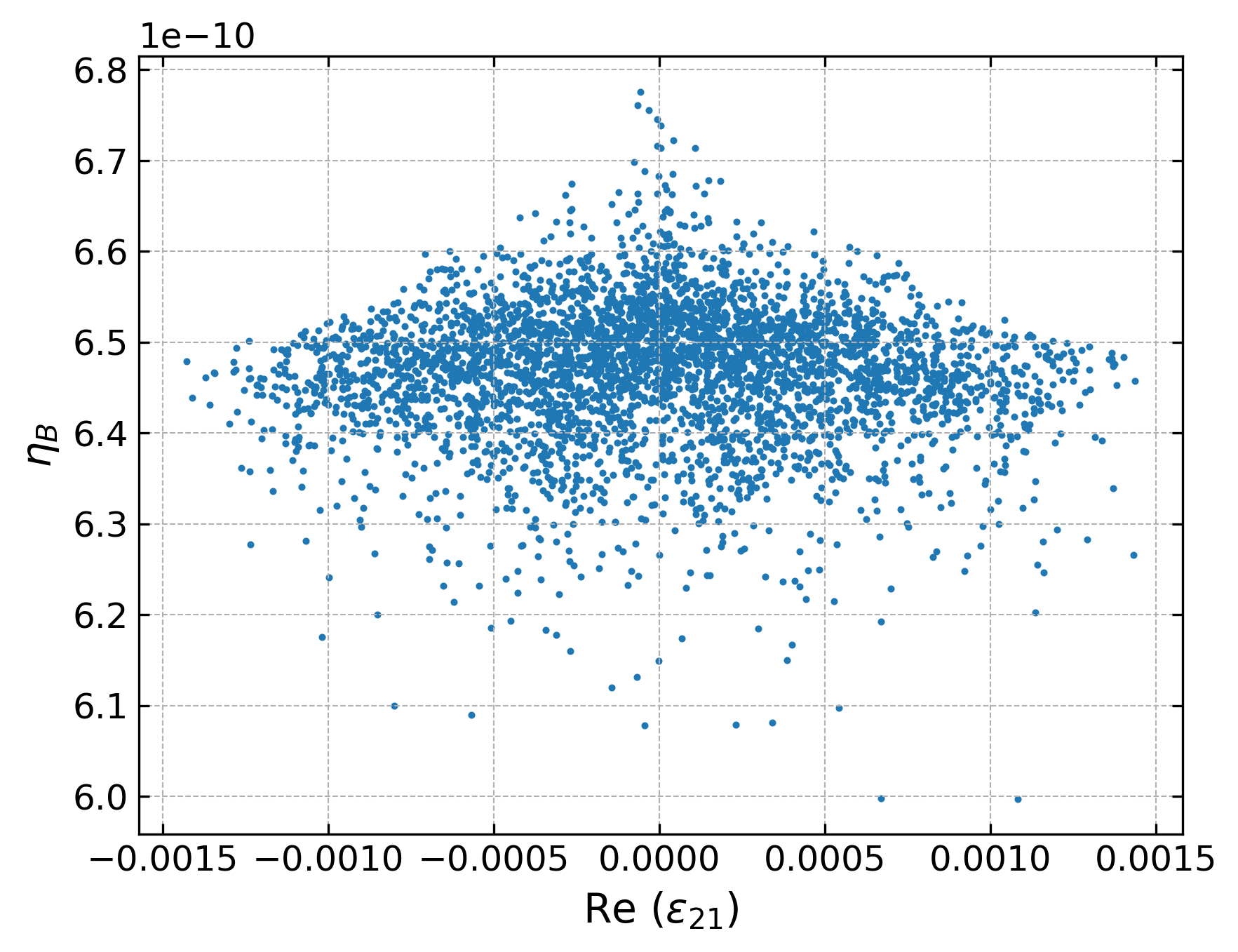}
     \end{subfigure}
     \hfill
     \begin{subfigure}[b]{0.46\textwidth}
         \centering
    \includegraphics[width=\textwidth]{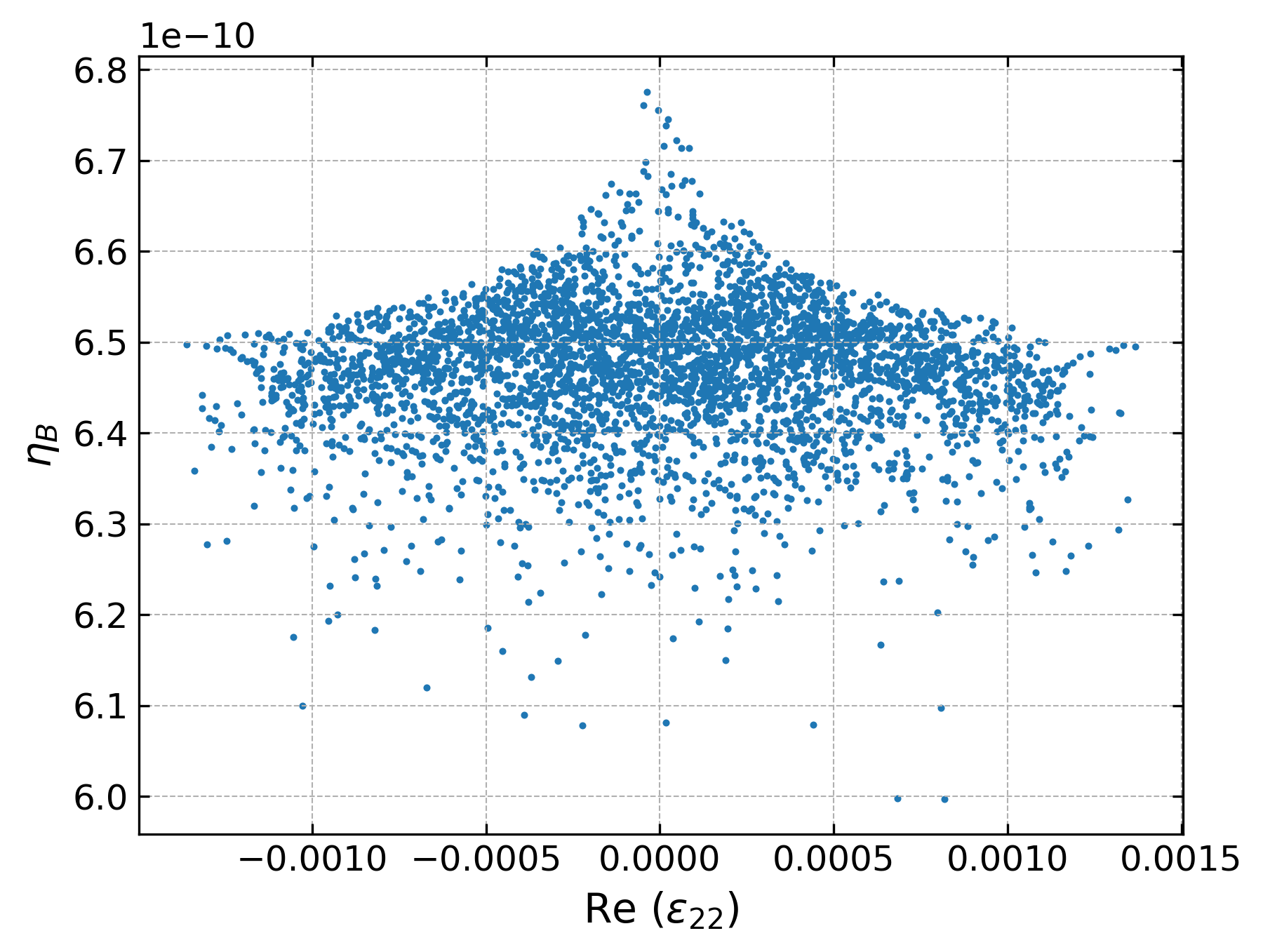}
     \end{subfigure}
  \caption{Correlations between the Baryon Asymmetry ($\eta_B$) with flavor-dependent CP-violating asymmetries $\varepsilon_{11}$ , $\varepsilon_{12}$, $\varepsilon_{21}$ and $\varepsilon_{22}$ respectively.}
  \label{fig5}
  \end{figure}

We illustrated that our resonant lepto-genesis scenario works well. We have the correlation of Baryon Asymmetry of the Universe with the flavor dependent CP-violating asymmetries in Fig. \ref{fig5}.

\section{Conclusion}
\label{conc}

We showed the $\Delta(54)$ flavor symmetry with SM Higgs boson and $Z_2 \otimes Z_3 \otimes Z_4$ symmetry which generates a neutrino mass matrix. Our model incorporates the ISS mechanism in order to provide a flavor-symmetric approach. This includes accounting for the solar mixing angle ,  upper octant of atmospheric mixing angle , non-zero reactor angle and CP violation ($\delta_{CP}$).  The values predicted for the absolute neutrino masses  are within the range of the cosmological bound  $\sum_{i} m_i  < 0.12$ eV. Neutrino oscillation parameter predictions based on the resulting mass matrix agree with the best-fit values obtained via $\chi^2$ analysis.  However, the Inverted Hierarchy (IH) scenario's predicted mixing angles, mass-squared disparities, and CP violation phase disagree with experimental findings.
 \par

Furthermore, we investigated baryogenesis via flavoured resonant leptogenesis. The right-handed neutrinos are degenerate at the dimension 5 level, introducing a higher dimension term resulted in a tiny splitting. We have taken
the splitting parameter, $d \approx 10^{-8}$ and thus, obtained a nonzero,
resonantly enhanced CP asymmetry ($\epsilon_{i\alpha}$) from the out-of equilibrium decay of right-handed Majorana neutrinos. We determined the baryon-to-photon ratio ($\eta_B$) in the flavored resonant leptogenesis scenario using the values of the CP asymmetries.  
 It was found that the model can explain
the observed value of BAU with particular choice of RHN mass scale, $M_1 = 10 TeV$ and mass splitting, $d \approx 10^{-8}$.

 Future work is reserved for examining the model to explore phenomena such as  Asymmetric Dark Matter.

\section*{Acknowledgements} \par

HB acknowledges Tezpur University, India for Institutional Research Fellowship. The research of NKF is funded by DST-SERB, India under Grant no. EMR/2015/001683.

\bibliographystyle{naturemag} 
\bibliography{references}

\begin{thebibliography}{10}
\expandafter\ifx\csname url\endcsname\relax
  \def\url#1{\texttt{#1}}\fi
\expandafter\ifx\csname urlprefix\endcsname\relax\def\urlprefix{URL }\fi
\providecommand{\bibinfo}[2]{#2}
\providecommand{\eprint}[2][]{\url{#2}}

\bibitem{aker2019improved}
\bibinfo{author}{Aker, M.} \emph{et~al.}
\newblock \bibinfo{title}{Improved upper limit on the neutrino mass from a direct kinematic method by katrin}.
\newblock \emph{\bibinfo{journal}{Physical review letters}} \textbf{\bibinfo{volume}{123}}, \bibinfo{pages}{221802} (\bibinfo{year}{2019}).

\bibitem{faessler2020status}
\bibinfo{author}{Faessler, A.}
\newblock \bibinfo{title}{Status of the determination of the electron--neutrino mass}.
\newblock \emph{\bibinfo{journal}{Progress in Particle and Nuclear Physics}} \textbf{\bibinfo{volume}{113}}, \bibinfo{pages}{103789} (\bibinfo{year}{2020}).

\bibitem{araki2005measurement}
\bibinfo{author}{Araki, T.} \emph{et~al.}
\newblock \bibinfo{title}{Measurement of neutrino oscillation with kamland: Evidence of spectral distortion}.
\newblock \emph{\bibinfo{journal}{Physical review letters}} \textbf{\bibinfo{volume}{94}}, \bibinfo{pages}{081801} (\bibinfo{year}{2005}).

\bibitem{cao2021physics}
\bibinfo{author}{Cao, S.} \emph{et~al.}
\newblock \bibinfo{title}{{Physics potential of the combined sensitivity of T2K-II, NO $\nu$ A extension, and JUNO}}.
\newblock \emph{\bibinfo{journal}{Physical Review D}} \textbf{\bibinfo{volume}{103}}, \bibinfo{pages}{112010} (\bibinfo{year}{2021}).

\bibitem{nath2021detection}
\bibinfo{author}{Nath, A.} \& \bibinfo{author}{Francis, N.~K.}
\newblock \bibinfo{title}{Detection techniques and investigation of different neutrino experiments}.
\newblock \emph{\bibinfo{journal}{International Journal of Modern Physics A}} \textbf{\bibinfo{volume}{36}}, \bibinfo{pages}{2130008} (\bibinfo{year}{2021}).

\bibitem{minkowski1977mu}
\bibinfo{author}{Minkowski, P.}
\newblock \bibinfo{title}{{$\mu$→ e$\gamma$ at a rate of one out of 109 muon decays?}}
\newblock \emph{\bibinfo{journal}{Physics Letters B}} \textbf{\bibinfo{volume}{67}}, \bibinfo{pages}{421--428} (\bibinfo{year}{1977}).

\bibitem{king2005testing}
\bibinfo{author}{King, S.} \& \bibinfo{author}{Yanagida, T.}
\newblock \bibinfo{title}{Testing the see-saw mechanism at collider energies}.
\newblock \emph{\bibinfo{journal}{Progress of theoretical physics}} \textbf{\bibinfo{volume}{114}}, \bibinfo{pages}{1035--1043} (\bibinfo{year}{2005}).

\bibitem{mohapatra1986mechanism}
\bibinfo{author}{Mohapatra, R.~N.}
\newblock \bibinfo{title}{Mechanism for understanding small neutrino mass in superstring theories}.
\newblock \emph{\bibinfo{journal}{Physical review letters}} \textbf{\bibinfo{volume}{56}}, \bibinfo{pages}{561} (\bibinfo{year}{1986}).

\bibitem{ma2009radiative}
\bibinfo{author}{Ma, E.}
\newblock \bibinfo{title}{Radiative inverse seesaw mechanism for nonzero neutrino mass}.
\newblock \emph{\bibinfo{journal}{Physical Review D}} \textbf{\bibinfo{volume}{80}}, \bibinfo{pages}{013013} (\bibinfo{year}{2009}).

\bibitem{mohapatra1999neutrino}
\bibinfo{author}{Mohapatra, R.~N.}, \bibinfo{author}{Nandi, S.} \& \bibinfo{author}{Perez-Lorenzana, A.}
\newblock \bibinfo{title}{Neutrino masses and oscillations in models with large extra dimensions}.
\newblock \emph{\bibinfo{journal}{Physics Letters B}} \textbf{\bibinfo{volume}{466}}, \bibinfo{pages}{115--121} (\bibinfo{year}{1999}).

\bibitem{arkani2001neutrino}
\bibinfo{author}{Arkani-Hamed, N.}, \bibinfo{author}{Dimopoulos, S.}, \bibinfo{author}{Dvali, G.} \& \bibinfo{author}{March-Russell, J.}
\newblock \bibinfo{title}{Neutrino masses from large extra dimensions}.
\newblock \emph{\bibinfo{journal}{Physical Review D}} \textbf{\bibinfo{volume}{65}}, \bibinfo{pages}{024032} (\bibinfo{year}{2001}).

\bibitem{fukugita1986barygenesis}
\bibinfo{author}{Fukugita, M.} \& \bibinfo{author}{Yanagida, T.}
\newblock \bibinfo{title}{Barygenesis without grand unification}.
\newblock \emph{\bibinfo{journal}{Physics Letters B}} \textbf{\bibinfo{volume}{174}}, \bibinfo{pages}{45--47} (\bibinfo{year}{1986}).

\bibitem{Nguyen:2018rlb}
\bibinfo{author}{Nguyen, T.~P.}, \bibinfo{author}{Le, T.~T.}, \bibinfo{author}{Hong, T.~T.} \& \bibinfo{author}{Hue, L.~T.}
\newblock \bibinfo{title}{{Decay of standard model-like Higgs boson $h\rightarrow \mu\tau$ in a 3-3-1 model with inverse seesaw neutrino masses}}.
\newblock \emph{\bibinfo{journal}{Phys. Rev. D}} \textbf{\bibinfo{volume}{97}}, \bibinfo{pages}{073003} (\bibinfo{year}{2018}).
\newblock \eprint{1802.00429}.

\bibitem{King:2014nza}
\bibinfo{author}{King, S.~F.}, \bibinfo{author}{Merle, A.}, \bibinfo{author}{Morisi, S.}, \bibinfo{author}{Shimizu, Y.} \& \bibinfo{author}{Tanimoto, M.}
\newblock \bibinfo{title}{{Neutrino Mass and Mixing: from Theory to Experiment}}.
\newblock \emph{\bibinfo{journal}{New J. Phys.}} \textbf{\bibinfo{volume}{16}}, \bibinfo{pages}{045018} (\bibinfo{year}{2014}).
\newblock \eprint{1402.4271}.

\bibitem{King:2003jb}
\bibinfo{author}{King, S.~F.}
\newblock \bibinfo{title}{{Neutrino mass models}}.
\newblock \emph{\bibinfo{journal}{Rept. Prog. Phys.}} \textbf{\bibinfo{volume}{67}}, \bibinfo{pages}{107--158} (\bibinfo{year}{2004}).
\newblock \eprint{hep-ph/0310204}.

\bibitem{Cao:2020ans}
\bibinfo{author}{Cao, S.} \emph{et~al.}
\newblock \bibinfo{title}{{Physics potential of the combined sensitivity of T2K-II, NO$\nu$A extension, and JUNO}}.
\newblock \emph{\bibinfo{journal}{Phys. Rev. D}} \textbf{\bibinfo{volume}{103}}, \bibinfo{pages}{112010} (\bibinfo{year}{2021}).
\newblock \eprint{2009.08585}.

\bibitem{Ahn:2014zja}
\bibinfo{author}{Ahn, Y.~H.} \& \bibinfo{author}{Gondolo, P.}
\newblock \bibinfo{title}{{Towards a realistic model of quarks and leptons, leptonic $CP$ violation, and neutrinoless $\beta\beta$-decay}}.
\newblock \emph{\bibinfo{journal}{Phys. Rev. D}} \textbf{\bibinfo{volume}{91}}, \bibinfo{pages}{013007} (\bibinfo{year}{2015}).
\newblock \eprint{1402.0150}.

\bibitem{mcdonald2016nobel}
\bibinfo{author}{McDonald, A.~B.}
\newblock \bibinfo{title}{Nobel lecture: the sudbury neutrino observatory: observation of flavor change for solar neutrinos}.
\newblock \emph{\bibinfo{journal}{Reviews of Modern Physics}} \textbf{\bibinfo{volume}{88}}, \bibinfo{pages}{030502} (\bibinfo{year}{2016}).

\bibitem{Nguyen:2020ehj}
\bibinfo{author}{Nguyen, T.~P.}, \bibinfo{author}{Thuc, T.~T.}, \bibinfo{author}{Si, D.~T.}, \bibinfo{author}{Hong, T.~T.} \& \bibinfo{author}{Hue, L.~T.}
\newblock \bibinfo{title}{{Low-energy phenomena of the lepton sector in an A4 symmetry model with heavy inverse seesaw neutrinos}}.
\newblock \emph{\bibinfo{journal}{PTEP}} \textbf{\bibinfo{volume}{2022}}, \bibinfo{pages}{023B01} (\bibinfo{year}{2022}).
\newblock \eprint{2011.12181}.

\bibitem{Hong:2022xjg}
\bibinfo{author}{Hong, T.~T.}, \bibinfo{author}{Nha, N. H.~T.}, \bibinfo{author}{Nguyen, T.~P.}, \bibinfo{author}{Phuong, L. T.~T.} \& \bibinfo{author}{Hue, L.~T.}
\newblock \bibinfo{title}{{Decays h \textrightarrow{} eaeb, eb \textrightarrow{} ea\ensuremath{\gamma}, and (g \ensuremath{-} 2)e,\ensuremath{\mu} in a 3-3-1 model with inverse seesaw neutrinos}}.
\newblock \emph{\bibinfo{journal}{PTEP}} \textbf{\bibinfo{volume}{2022}}, \bibinfo{pages}{093B05} (\bibinfo{year}{2022}).
\newblock \eprint{2206.08028}.

\bibitem{kajita2016nobel}
\bibinfo{author}{Kajita, T.}
\newblock \bibinfo{title}{Nobel lecture: Discovery of atmospheric neutrino oscillations}.
\newblock \emph{\bibinfo{journal}{Reviews of Modern Physics}} \textbf{\bibinfo{volume}{88}}, \bibinfo{pages}{030501} (\bibinfo{year}{2016}).

\bibitem{de20212020}
\bibinfo{author}{de~Salas, P.~F.} \emph{et~al.}
\newblock \bibinfo{title}{2020 global reassessment of the neutrino oscillation picture}.
\newblock \emph{\bibinfo{journal}{Journal of High Energy Physics}} \textbf{\bibinfo{volume}{2021}}, \bibinfo{pages}{1--36} (\bibinfo{year}{2021}).

\bibitem{okada2021spontaneous}
\bibinfo{author}{Okada, H.} \& \bibinfo{author}{Tanimoto, M.}
\newblock \bibinfo{title}{{Spontaneous CP violation by modulus $\tau$ in A4 model of lepton flavors}}.
\newblock \emph{\bibinfo{journal}{Journal of High Energy Physics}} \textbf{\bibinfo{volume}{2021}}, \bibinfo{pages}{1--27} (\bibinfo{year}{2021}).

\bibitem{Ahn:2021ndu}
\bibinfo{author}{Ahn, Y.~H.}, \bibinfo{author}{Kang, S.~K.} \& \bibinfo{author}{Lee, H.~M.}
\newblock \bibinfo{title}{{Toward a model of quarks and leptons}}.
\newblock \emph{\bibinfo{journal}{Phys. Rev. D}} \textbf{\bibinfo{volume}{106}}, \bibinfo{pages}{075029} (\bibinfo{year}{2022}).
\newblock \eprint{2112.13392}.

\bibitem{PhongNguyen:2017meq}
\bibinfo{author}{Phong~Nguyen, T.}, \bibinfo{author}{Hue, L.~T.}, \bibinfo{author}{Si, D.~T.} \& \bibinfo{author}{Thuc, T.~T.}
\newblock \bibinfo{title}{{CP violations in a predictive $A_4$ symmetry model}}.
\newblock \emph{\bibinfo{journal}{PTEP}} \textbf{\bibinfo{volume}{2020}}, \bibinfo{pages}{033B04} (\bibinfo{year}{2020}).
\newblock \eprint{1711.05588}.

\bibitem{Buravov:2014dna}
\bibinfo{author}{Buravov, L.~I.}
\newblock \bibinfo{title}{{Confining potential and mass of elementary particles}}.
\newblock \emph{\bibinfo{journal}{J. Mod. Phys.}} \textbf{\bibinfo{volume}{7}}, \bibinfo{pages}{129--133} (\bibinfo{year}{2017}).
\newblock \eprint{1502.00958}.

\bibitem{buravov2009elementary}
\bibinfo{author}{Buravov, L.}
\newblock \bibinfo{title}{{Elementary muon, pion, and kaon particles as resonators for neutrino quanta. Calculations of mass ratios for e, $\mu$, $\pi$ 0, $\pi$$\pm$, K 0, K$\pm$, and $\nu$ e}}.
\newblock \emph{\bibinfo{journal}{Russian Physics Journal}} \textbf{\bibinfo{volume}{52}}, \bibinfo{pages}{25--32} (\bibinfo{year}{2009}).

\bibitem{barman2023neutrino}
\bibinfo{author}{Barman, A.}, \bibinfo{author}{Francis, N.~K.} \& \bibinfo{author}{Bora, H.}
\newblock \bibinfo{title}{{Neutrino Mixing Phenomenology: $A_4$ Discrete Flavor Symmetry with Type-I Seesaw Mechanism}}.
\newblock \emph{\bibinfo{journal}{arXiv preprint arXiv:2306.11461}}  (\bibinfo{year}{2023}).

\bibitem{bora2023majorana}
\bibinfo{author}{Bora, H.}, \bibinfo{author}{Francis, N.~K.}, \bibinfo{author}{Barman, A.} \& \bibinfo{author}{Thapa, B.}
\newblock \bibinfo{title}{{Majorana neutrinos in Inverse Seesaw and $\Delta (54)$ Flavor Models}}.
\newblock \emph{\bibinfo{journal}{arXiv e-prints}} \bibinfo{pages}{arXiv--2311} (\bibinfo{year}{2023}).

\bibitem{kuzmin1985anomalous}
\bibinfo{author}{Kuzmin, V.~A.}, \bibinfo{author}{Rubakov, V.~A.} \& \bibinfo{author}{Shaposhnikov, M.~E.}
\newblock \bibinfo{title}{On anomalous electroweak baryon-number non-conservation in the early universe}.
\newblock \emph{\bibinfo{journal}{Physics Letters B}} \textbf{\bibinfo{volume}{155}}, \bibinfo{pages}{36--42} (\bibinfo{year}{1985}).

\bibitem{davidson2002lower}
\bibinfo{author}{Davidson, S.} \& \bibinfo{author}{Ibarra, A.}
\newblock \bibinfo{title}{A lower bound on the right-handed neutrino mass from leptogenesis}.
\newblock \emph{\bibinfo{journal}{Physics Letters B}} \textbf{\bibinfo{volume}{535}}, \bibinfo{pages}{25--32} (\bibinfo{year}{2002}).

\bibitem{fong2021low}
\bibinfo{author}{Fong, C.~S.}, \bibinfo{author}{Rahat, M.~H.} \& \bibinfo{author}{Saad, S.}
\newblock \bibinfo{title}{{Low-scale resonant leptogenesis in SU(5) GUT with T13 family symmetry}}.
\newblock \emph{\bibinfo{journal}{Phys. Rev. D}} \textbf{\bibinfo{volume}{104}}, \bibinfo{pages}{095028} (\bibinfo{year}{2021}).
\newblock \eprint{2103.14691}.

\bibitem{bora2024neutrino}
\bibinfo{author}{Bora, H.}, \bibinfo{author}{Francis, N.~K.}, \bibinfo{author}{Barman, A.} \& \bibinfo{author}{Thapa, B.}
\newblock \bibinfo{title}{{Neutrino mass model in the context of $ \Delta(54)\otimes Z_2 \otimes Z_3 \otimes Z_4$ flavor symmetries with Inverse Seesaw mechanism}}.
\newblock \emph{\bibinfo{journal}{Physics Letters B}} \textbf{\bibinfo{volume}{848}}, \bibinfo{pages}{138329} (\bibinfo{year}{2024}).

\bibitem{thapa2021resonant}
\bibinfo{author}{Thapa, B.} \& \bibinfo{author}{Francis, N.~K.}
\newblock \bibinfo{title}{{Resonant leptogenesis and \uppercase{TM}$_1$ mixing in minimal type-\uppercase{I} seesaw model with \uppercase{S}$_4$ symmetry}}.
\newblock \emph{\bibinfo{journal}{The European Physical Journal C}} \textbf{\bibinfo{volume}{81}}, \bibinfo{pages}{1--8} (\bibinfo{year}{2021}).

\bibitem{ishimori2010non}
\bibinfo{author}{Ishimori, H.} \emph{et~al.}
\newblock \bibinfo{title}{Non-abelian discrete symmetries in particle physics}.
\newblock \emph{\bibinfo{journal}{Progress of Theoretical Physics Supplement}} \textbf{\bibinfo{volume}{183}}, \bibinfo{pages}{1--163} (\bibinfo{year}{2010}).

\bibitem{ishimori2009lepton}
\bibinfo{author}{Ishimori, H.}, \bibinfo{author}{Kobayashi, T.}, \bibinfo{author}{Okada, H.}, \bibinfo{author}{Shimizu, Y.} \& \bibinfo{author}{Tanimoto, M.}
\newblock \bibinfo{title}{{Lepton flavor model from $\Delta$ (54) symmetry}}.
\newblock \emph{\bibinfo{journal}{Journal of High Energy Physics}} \textbf{\bibinfo{volume}{2009}}, \bibinfo{pages}{011} (\bibinfo{year}{2009}).

\bibitem{lei2020minimally}
\bibinfo{author}{Lei, M.} \& \bibinfo{author}{Wells, J.~D.}
\newblock \bibinfo{title}{{Minimally modified $A_4$ Altarelli-Feruglio model for neutrino masses and mixings and its experimental consequences}}.
\newblock \emph{\bibinfo{journal}{Physical Review D}} \textbf{\bibinfo{volume}{102}}, \bibinfo{pages}{016023} (\bibinfo{year}{2020}).

\bibitem{pilaftsis2004resonant}
\bibinfo{author}{Pilaftsis, A.} \& \bibinfo{author}{Underwood, T.~E.}
\newblock \bibinfo{title}{Resonant leptogenesis}.
\newblock \emph{\bibinfo{journal}{Nuclear Physics B}} \textbf{\bibinfo{volume}{692}}, \bibinfo{pages}{303--345} (\bibinfo{year}{2004}).

\bibitem{pilaftsis1997cp}
\bibinfo{author}{Pilaftsis, A.}
\newblock \bibinfo{title}{Cp violation and baryogenesis due to heavy majorana neutrinos}.
\newblock \emph{\bibinfo{journal}{Physical Review D}} \textbf{\bibinfo{volume}{56}}, \bibinfo{pages}{5431} (\bibinfo{year}{1997}).

\bibitem{xing2020bridging}
\bibinfo{author}{Xing, Z.-z.} \& \bibinfo{author}{Zhang, D.}
\newblock \bibinfo{title}{Bridging resonant leptogenesis and low-energy cp violation with an rge-modified seesaw relation}.
\newblock \emph{\bibinfo{journal}{Physics Letters B}} \textbf{\bibinfo{volume}{804}}, \bibinfo{pages}{135397} (\bibinfo{year}{2020}).

\end{thebibliography}

\end{document}